\documentclass[journal=jacsat,lineno,manuscript=article]{achemso}
\mciteErrorOnUnknownfalse 
\usepackage[version=3]{mhchem} 
\usepackage{sistyle} 
\usepackage{amsmath,amssymb,gensymb,amsfonts,latexsym,mathtools} 
\usepackage{empheq,xfrac} 

\usepackage[hidelinks,linktocpage]{hyperref} 

\usepackage{array,makecell} 
\usepackage{booktabs}	
\usepackage{multicol,multirow} 
\usepackage{dcolumn}  
\usepackage{xcolor}
\usepackage{soul}

\author{Jose M. Sojo-Gordillo}
\email{jose.sojo@unibas.ch}
\affiliation{University of Basel, Klingelbergstrasse 82, 4056, Basel, Switzerland}

\author{Ahmed Alwakil}
\affiliation{Univ Lyon, CNRS, INSA-Lyon, Université Claude Bernard Lyon 1, CETHIL UMR5008, F-69621, Villeurbanne, France}

\author{Olivier Merchiers}
\affiliation{Univ Lyon, CNRS, INSA-Lyon, Université Claude Bernard Lyon 1, CETHIL UMR5008, F-69621, Villeurbanne, France}

\author{Pierre-Olivier Chapuis}
\affiliation{Univ Lyon, CNRS, INSA-Lyon, Université Claude Bernard Lyon 1, CETHIL UMR5008, F-69621, Villeurbanne, France}
\email{pierre-olivier.chapuis@insa-lyon.fr}

\newcommand{\thetitle}{Thermal radiation of nanowires and near-field interaction with substrates: exact computation and proximity approximation}
\title[Manuscript]{\thetitle}

\abbreviations{NFRHT,NW}

\begin{document}

\begin{tocentry}
Radiative heat transfer involving nanowires is computed. Both emission by the sub-wavelength wire and near-field interaction with flat substrates are considered. Gold nanowires are particularly emissive, in contrast to silicon ones. Radiative exchange can exceed the Stefan-Boltzmann prediction by one order of magnitude when a substrate is brought in close proximity. Different levels of approximation are discussed.

\includegraphics[width=1\textwidth]{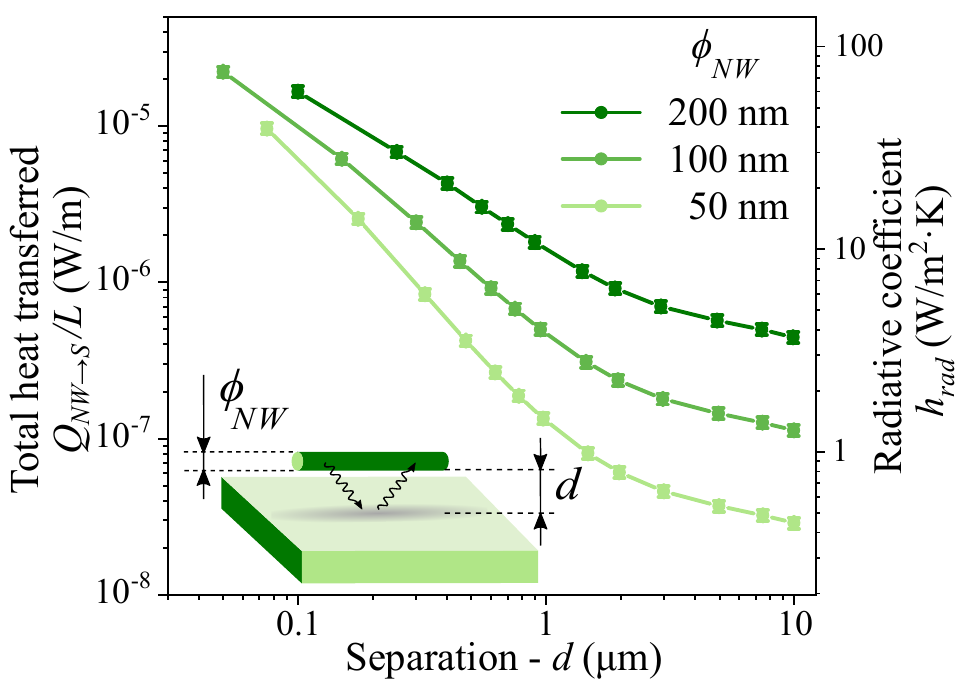}

\end{tocentry}

\clearpage
\begin{abstract}

Radiative heat transfer involving complex nano-objects is difficult to evaluate due to their sub-wavelength size, which induces deviation to macroscopic Planck's opaque surface-to-surface formalism, and due to the small distances involved, which lead to near-field radiative (photon tunneling) effects. As a result, thermal radiation has mostly been neglected in analyses of thermal transport in nanowires. Here, we report on calculations of thermal emission by single nanowires and on the radiative interaction between suspended silicon nanowires and flat substrates. We observe that metallic nanowires can be particularly emissive in comparison to wires made of other materials. While thermal emission from an isolated silicon nanowire is lower than Stefan-Boltzmann's prediction, the radiative exchange with a substrate can be larger by one order of magnitude. The considered configurations also allow benchmarking various levels of common or original approximations, including, in near field, the proximity approximation, the thickness-dependent PA, the electrostatic approach, and the effective emissivity in the far field.

\end{abstract}


\newpage

Classical (Planckian) theory of surface-to-surface radiative transfer between opaque bodies only holds for macroscopic sizes. Because thermal-radiation rays -- electromagnetic (EM) waves -- are screened as they travel inside media, only a skin layer close to the surfaces of a body contributes to the associated radiative heat transfer. However, at the micro and nanoscale, sizes become comparable to the skin-layer one and considering thermal-radiation as a surface effect is not anymore relevant \cite{Planck1914}. In particular, the emitted energy distribution given by Planck's law no longer holds if characteristic sizes are smaller than the thermal wavelength $\lambda_W T$ ($\lambda_W T$ = \SI{2890}{\micro m.K} so $\lambda_W \sim$ \SI{10}{\micro m} at room temperature). Hence, for bodies smaller than this size, the concept of Sub-Wavelength Thermal Emission (SWTE) must be addressed. In this case, the emissivity concept associated with the surface theory -- with its upper bound equal to unity \textit{i.e.} the black-body limit -- breaks down. This was shown theoretically \cite{Kattawar1970, Krueger2012,Nguyen2018} and more-recently experimentally \cite{Thompson2020,Shin2019}. Moreover, when distances between bodies exchanging thermal radiation also become smaller than $\lambda_W$, tunneling of evanescent waves becomes possible, locally boosting the heat transfer rate between them. This effect is known as Near-Field Radiative Heat Transfer (NFRHT) \cite{Polder1971}, and is now well-documented both numerically \cite{Zeng2025} and experimentally \cite{Lucchesi2021}. It is worth noticing that both phenomena are not mutually exclusionary, since nanometric bodies, i.e. sub-wavelength-sized ones, can also experiment NFRHT with other bodies in proximity. In the following, exact radiative heat transfer rates involving the nano-objects and sizes will be compared with those predicted by (mis)using the macroscopic blackbody theory, whose orders of magnitude and typical features are well known and therefore provide references for analyses, exhibiting super-Planckian performances\cite{Guo2012}. We will discuss various levels of approximations of the full theory (fluctuational electrodynamics, FE) \cite{Rytov1959}. Geometrical complexity is not always easy to deal (either analytically or numerically), and concepts such as proximity approximation (PA)\cite{Derjaguin1956} and electrostatic approximation (EA) in the near field or effective emissivity (EE) in the far field will be discussed as means to simplify the estimation of the radiative flux. In particular, the PA will be extended by considering finite thickness of the emitting body.

More specifically, this work focuses on the radiative exchanges involving suspended nanowires. A huge body of works has dealt with these objects in the last 20 years, in particular with silicon-based ones, demonstrating how heat conduction is modified by phonon confinement and the rough nature of the wire surfaces \cite{Li2003, Chen2008, Lim2012, Lee2016, Yang2021, Sojo2024}. In most of these works, thermal radiation was simply neglected, either based on erroneous macroscopic arguments or simply because it was not possible to evaluate it. One aim of this work is to 
provide hints for evaluating thermal radiation correctly in many of these nanowire-based experiments. Therefore, two configurations are considered: either a nanowire remote from any substrate, where thermal emission by the nanowire alone is to be accounted for, or a nanowire lying close to a substrate, as it is often the case\cite{Gadea2019, Sojo2022, Sojo2023}. Results are also reported as heat exchange coefficients for convenience. The wire radial temperature profiles are considered homogeneous if their Biot number ($Bi=h_{rad} \phi/\kappa$, where $\kappa$ is the thermal conductivity, $h_{rad}$ the equivalent radiative heat transnfer coefficient, and $\phi$ the nanowire diameter) stays small, such assumption will be verified in the following.



Radiative heat transferred from body $i$ to $j$ can be expressed in a Landauer-like formula as\cite{Biehs2021,Krueger2012}:

    \begin{equation}
        Q_{i \rightarrow j} = \int_{0}^{\infty} \left[\Theta \left(T_i, \omega\right) - \Theta \left(T_j, \omega\right) \right] \cdot \Psi(\omega) d\omega,
            \label{eq:HeatTransferred}
    \end{equation} 

where $T_i$ is the temperature of the body $i$, $\Psi(\omega)$ is a transmission function describing the frequency-dependent fluctuations in body $j$ produced by $i$ in the far or near field, and $\Theta \left(T_i, \omega\right) = \hbar \omega / \left( \exp \left( \frac{\hbar \omega}{k_B T_i} \right) - 1 \right) $ is the Bose-Einstein factor, which describes the mean energy of a Planck oscillator at the frequency $\omega$ in thermal equilibrium at temperature $T$. Therefore, the problem resides in accurately describing the transmission function $\Psi(\omega)$, which accounts for all the geometrical aspects as well as for the dielectric and magnetic (optical) properties of the emitters, receivers, and transfer media. Here, we neglect the temperature dependence of the optical properties and therefore their impact on $\Psi$ (see e.g. \cite{Nguyen2018} for this issue). Hence, aside from the geometrical description of the problem, the sole parameter needed to perform the simulation is a function that describes the complex refractive index $n$, or equivalently, the complex permittivity $\epsilon = n^2$ for each of the media considered. 

\begin{figure}[tb]
\begin{center}
    \includegraphics[width=0.6\textwidth]{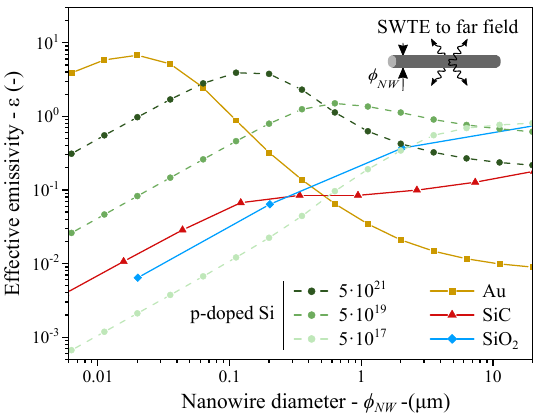}
      \caption{Effective emissivity, i.e. radiative power emitter per unit area, of cylindrical wires as a function of their diameters for various materials.}
      \label{fig:EpsilonVsRadii}
\end{center}
\end{figure}

First, the impact of size on thermal-radiation emission by cylindrical wires is analyzed, ranging from thin nanowires up to microwires. Analytical expressions described by \citet{Golyk2012}, \citet{Xiao2023}, and \citet{Alwakil} were used to compute the emitter power. Results were obtained for typical materials, including a metal (\ce{Au}), amorphous (\ce{SiO2}), crystalline (\ce{SiC}) dielectrics, and silicon with increasing doping levels ranging from \SI{5e17}{} to \SI{5e21}{cm^{-3}}. The complex permittivity $\epsilon$ of heavily doped (degenerated) silicon ($N > $~\SI{5e17}{cm^{-3}}) and gold was calculated using the Drude model\cite{Park2013}. Here, the effective emissivity $\varepsilon$ \cite{Nguyen2017} (not to be confused with the dielectric function $\epsilon = n^2$) is defined as the ratio of the emitted power divided by the product of the area and $\sigma T^4$, where $\sigma$ is the Stefan-Boltzmann constant. If the wire is surrounded by a large medium close to the wire temperature, the radiative exchange coefficient, defined as

\begin{equation} \label{eq:hrad}
    h_{rad} = \frac{Q_{NW \rightarrow \infty} \cdot L} {A \cdot \Delta T} = \frac{Q_{NW \rightarrow \infty}} {\pi \phi \Delta T},
\end{equation}

where $Q_{NW \rightarrow \infty}$ is the sub-wavelength thermal emission (SWTE) radiated towards the far field surrounding,  $\phi$ is the wire diameter, and $L$ its length, can be computed as $h_{rad}=\varepsilon 4 \sigma T^3$. \autoref{fig:EpsilonVsRadii} shows that effective emissivity possesses a maximum that exceeds unity for materials such as gold with negative real permittivity. This feature was already observed for dielectric materials in spherical particles \cite{Nguyen2017}, but not for metallic ones. Note that larger effective emissivity was predicted for strongly-anisotropic bodies \cite{FernandezHurtado2018}. Interestingly, while gold is usually known as a bad emitter, gold nanowires appear to emit much more than their counterparts made of other materials (note that we have not considered electron confinement here, which could further enhance emission). This may be linked to emission of gold planar thin film, which increases when thickness decreases \cite{Edalatpour2013} but stays bounded by unity. This confirms that typical behaviours at macroscopic scale are not necessarily reproduced at nanoscale. For the case of silicon with increasing levels of doping, emissivity increases proportionally with the doping level, yielding intermediate emissivities between dielectrics such as \ce{SiO2} or \ce{SiC} and pure metals. As the doping level increases, the maxima also shifts to lower radii, trending towards pure metal behavior. This trend is related to the ratio of wire diameter to penetration depth $\delta$, which is close to unity for radii of \SI{10}{nm} in the case of \ce{Au}.

\begin{figure}[!tb]
\begin{center}
    \includegraphics[width=0.6\textwidth]{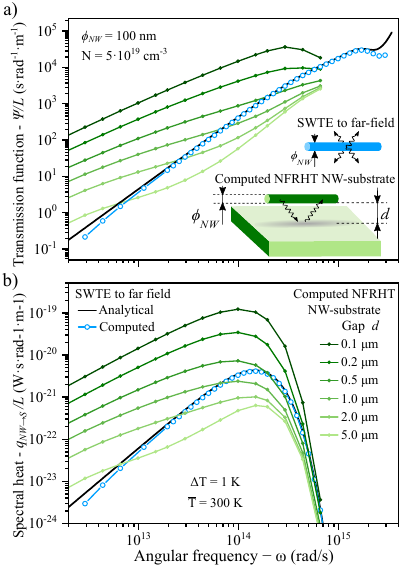}
      \caption{a)~Calculated transmission functions per unit length $\Psi/L$ for a \SI{100}{nm} in diameter NW and a doping level of \SI{5e19}{cm^{-3}}. The solid black line represent the SWTE of a NW towards the far-field environment as sketched in the top inset diagram. The dotted blue line represent the calculated $\Psi/L$ function using the BEM model. Green dotted lines show the calculated NFRHT $\Psi/L$ for the NW parallel to a flat substrate at different $d$ gap distances as sketched in the bottom inset diagram. b)~Computed spectral heat transfer per unit length from nanowire to substrate $q_{NW \rightarrow S}/L$ calculated for the same NW assuming a temperature difference of \SI{1}{K} at an average temperature of \SI{300}{K}. 
      }
      \label{fig:Spectral_Phi100_study}
\end{center}
\end{figure}  

For more complex geometries, numerical simulations based on the Boundary Element Method (BEM) can be used. The BEM calculations were implemented within the open-source code Scuff-EM\cite{Reid2015,Kim2015a,Nguyen2017}. Validation of the BEM model is performed by estimating SWTE of an (supposed) infinitely long cylinder toward a far-field environment (effectively acting as a black-body) as sketched in the upper inset of \autoref{fig:Spectral_Phi100_study}a and comparing it to the analytical solution\cite{Golyk2012}. For the simulated case, the emission of a \SI{10}{\micro m} long NWs of diameters ranging from \SI{50}{} to \SI{200}{nm} were computed and exemplified for the specific case of \SI{100}{nm} in \autoref{fig:Spectral_Phi100_study}. The NW was simulated with a doping level of \SI{5e19}{cm^{-3}}. As observed in \autoref{fig:Spectral_Phi100_study}a, the computed $\Psi/L$ values reproduced accurately the analytical model with relative errors of less than 12\% in the range of spectral emission (\SI{5e12}{} to \SI{2e15}{rad/s}). Deviations in the high-frequency range were found, likely as a consequence of surface discretization (see \cite{Nguyen2017} for constrains to be fulfilled). However, those mismatches have already a negligible influence on the total heat transferred since, at the temperature studied, the number of these high-frequency modes is null. Opposite, significantly lower transferred heat is found in the low-frequency range, related to the confinement of modes larger than the \SI{10}{\micro m} of the simulated cylinder. Indeed, additional simulations (not shown here) showed how increasing the length of the modeled cylinder improves the accuracy of results at this low-frequency range. Nevertheless, the contribution of this part of the spectrum to the total integrated heat is also negligible compared to intermediate frequencies (\SI{5e12}{} -- \SI{2e15}{rad/s}). Hence, the error in the total heat transfer derived from this mismatch is also negligible and thus not worth in terms of computational cost.

The radiative exchange between the NW and a flat surface was also computed. Heat transfer simulations (sketched in the bottom schematic of \autoref{fig:Spectral_Phi100_study}a) were performed using discretizations with a minimum number of $N_{facets}$ of $10^4$ for a cylinder of similar length (\SI{10}{\micro m}), which ensured deviations from the exact solution within a 10\% error margin for the case of SWTE. Both solids (cylinder and bulk box) were simulated at the same doping level (\SI{5e19}{cm^{-3}}) and therefore show the same permitivity function.

\autoref{fig:Spectral_Phi100_study} describes the spectral transmission function $\Psi/L$ (a) and the spectral heat transfer per unit length $q_{NW \rightarrow S}/L$ (b) between a NW of \SI{100}{nm} in diameter and a parallel surface separated by distances ranging from 0.1 to \SI{5}{\micro m}. $q_{NW \rightarrow S}/L$ presents a maximum at \SI{3e14}{rad/s} for the shortest separation, corresponding with the strongest interaction of the NFRHT effect. This region is the most significant contribution of the total transferred heat. Moreover, it can also be appreciated how, for separation gaps larger than \SI{1}{\micro m}, all $q_{NW \rightarrow S}/L$ converged to the same values at the high frequency range ($\omega >$ \SI{1.5e14}{rad/s}). This suggests that those separation distances are larger than the range of near-field interaction at those specific frequencies. Indeed, compared to the SWTE values, this interaction is weaker for most part of the spectra, indicating that this distance defines the threshold below which NFRHT becomes the dominant mechanism. 

\begin{figure}[!tb]
\begin{center}
\includegraphics[width=0.6\textwidth]{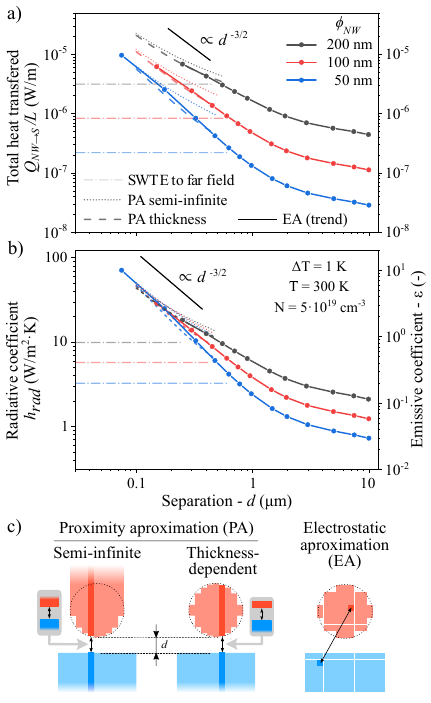}
\caption{a)~Total transferred heat per unit length from nanowire to substrate $Q_{NW \rightarrow S}/L$ as a function of the distance for three different nanowire diameter considering a temperature difference $\Delta T$ of \SI{1}{K} at an average temperature of \SI{300}{K} (solid lines with circles). The dashed and dotted lines represent the two proximity approximations added to the view-factor macroscopic estimation considering the SWTE effective emissivity calculated for the nanowire. Dash-dot lines illustrate the SWTE of the standalone NWs as a reference. b)~Estimated radiative coefficient $h_{rad}$ calculated using \autoref{eq:hrad}. The right scale shows the equivalent effective emissivity of the transfer. c)~Schematic of the different levels of approximation of \autoref{eq:Aproximations}. }
\label{fig:TotQ_hrad_Dstudy}
\end{center}
\end{figure}  

\autoref{fig:TotQ_hrad_Dstudy}a shows the total transferred heat between nanowire and surface as a function of the distance for three different nanowire diameters. Close to the surface (at $d<$\SI{1}{\micro m}) the heat transfer scales with distance as $d^{-1.4}$. The cylinder with \SI{200}{nm} diameter showed the highest heat transfer. Reduced diameters tailored the total heat transfer. Additionally, \autoref{fig:TotQ_hrad_Dstudy}b illustrates the computed radiative coefficient $h_{rad}$ (see \autoref{eq:hrad}) as a function of the separation distance for the three nanowire radii studied. As depicted by the solid lines, a strong increase of the coefficient $h_{rad}$ with respect to values predicted by classical blackbody theory is predicted, especially for small radii (remind that $h_{rad} \approx 4 \varepsilon \sigma T^3 \approx 6.12 $~$\varepsilon $~  W.m$^{-2}$.K$^{-1}$ for an object slightly heated/cooled respect to the medium in which it is immersed). Combining the range of $h_{rad}$ obtained with experimental thermal conductivities of NWs within the considered diameter and doping range (\SI{5}{}-\SI{25}{W/m.K})\cite{Li2003,Lim2012,Lee2016,Gadea2019,Sojo2022} to estimate the Biot number, we can conclude that temperature stays radially homogeneous.

Finally, it is interesting to compare these results with various levels of approximations (see \autoref{fig:TotQ_hrad_Dstudy}c) allowing to bypass the difficulty of the geometrical complexity. It has been customary to apply the proximity approximation (PA) to various kinds of geometries. While initially grounded on the argumentation that thermal-radiation is due to the strong distance-dependent dipole-dipole interaction ($d^{-6}$), it was shown that its range of validity is sometimes extended \cite{Clayton2011,Karthik2011,Lucchesi2021}. PA considers all objects as locally parallel, and power exchange rates can be obtained from a power-distance curve for semi-infinite flat parallel media previously generated once and used for all cases. As a result, it disentangles the geometry and electromagnetic problems. Here, the nanowire diameters considered are thin, so extrapolating such calculation is questionable. Indeed, if the skin depth $\delta$ is larger than the object size, the PA will overestimate the volumes that interact. We therefore introduce the thickness-dependent proximity approximation (TD-PA, see \autoref{fig:TotQ_hrad_Dstudy}c), which is based on the initial tabulation of the flux between thin films of varying thicknesses and planar semi-infinite media. This is aimed to include the effect of reduced thickness for the emission. Yet, it still neglects local curvature. 

Another level of analysis can be performed with the electrostatic approximation (EA). It is correct at very small distances, when retardation becomes negligible. For thin nanowires close to substrate, it could therefore be accurate. EA is also known to provide the asymptotics of non-metallic bodies at small distances. Integrating the $d^{-6}$ dependence of the power exchange between local dipoles, we find the radiative exchange coefficient writing as

    \begin{equation}
        Q_{i \rightarrow j}^{NF} \propto \frac{\pi R^2}{d^{3/2}(d+2R)^{3/2}}.
            \label{eq:Aproximations}
    \end{equation} 
Depending on the ratio $R/d$ (providing both stay smaller than the wavelength), one finds asymptotic behaviours as $\pi R^2/d^3$ (like for the point dipole \cite{Mulet2001}) or a $\sqrt{R/d^3}$.

All these approximations (PA, TD-PA, EA) target the near-field zone, i.e. where $d < \lambda_W$. Therefore, it can be wise to include another type of approximation for the far-field contribution to the total flux, as applying the PA is not relevant anymore in this zone\cite{Lucchesi2021}. For far field, view-factor theory may still be valid, provided one considers the effective emissivity (EE) of the sub-wavelength body. The different near-field approximations are depicted in \autoref{fig:TotQ_hrad_Dstudy}.

The main results are as follows: (i)~all curves tend to the asymptotics given by EA, (ii)~while the PA indeed overestimates the exchanged flux, the TD-PA seems to underestimate it, (iii)~the EE does not seem to work when total emissivities are considered (the result towards far field result is overestimated by a factor 3, and thus not shown). The last observation questions the practical use of the EE concept in complex exchanges. In general, these results underline the difficulty in applying these approximations over a large range of distance in the case of complex objects exchanging heat by means of thermal radiation.


The previous simulations showed how the studied heavily doped silicon nanowires with sub-wavelength sizes present effective emissivities higher than unity. Moreover, for more than two bodies spaced by sub-wavelength distances, the heat exchange can get enhanced by an additional order of magnitude due to NFRHT. Both cases represent super-Planckian examples beyond the prediction from the blackbody macroscopic theory. However, it is worth noticing that the case here presented was simulated assuming the same doping concentration as for the NW ($N =$ \SI{5e19}{cm^{-3}}). Intrinsic Si will have different behavior as the penetration depth $\delta$ grows significantly in the frequency range simulated. This would require simulating larger sections of bulk, representing an implementation challenge that is currently being investigated. 

In summary, the obtained radiative exchange coefficients are significantly larger than those predicted by classical blackbody radiation theory for distances below the thermal wavelength ($\lambda_W \sim$ \SI{10}{\micro m}). Hence, both sub-wavelength thermal emission and near-field radiative heat transfer need to the accounted if an accurate evaluation of the thermal properties of nanowires suspended over substrates at closer distances is intended. Thickness-dependent proximity approximation provides a reasonably good estimation of the heat transferred between the two objects for gaps below \SI{500}{nm}, whereas further refinement of the models is required to match the behavior in the transition regime to far field. 

\begin{acknowledgement}

This investigation was supported by the Spanish Ministry of Education through the FPU grant \textbf{FPU18/01494}. The European Commission supported this investigation through the \textit{EFINED} project \textbf{H2020-FETOPEN-2016-2017 GA 766853} and the French Research Agency ANR through the \textit{CASTEX} project \textbf{ANR-21-CE30-0027}. 

\end{acknowledgement}

\section{Author contributions}
\textbf{J.M.S.G.:} conceptualization, investigation, data curation, formal analysis, visualization, software, writing - original draft;
\textbf{A.A.:} investigation, data curation, software;
\textbf{O.M.:} software, investigation, data curation, formal analysis, software, writing - original draft;
\textbf{P.O.C.:} conceptualization, investigation, data curation, formal analysis, resources, funding acquisition, writing - original draft;



 \small{\bibliography{References}}
 \clearpage


    
\end{document}